\documentclass[a4paper,11pt]{article}
\usepackage[utf8]{inputenc}
\usepackage{lmodern}
\usepackage[T1]{fontenc}
\usepackage{amsmath,amsthm,amsfonts,amssymb,bm}
\usepackage{dsfont}     % indicator function
\usepackage{enumitem}
\usepackage{graphicx}
\usepackage{booktabs}
\usepackage{xcolor}
\usepackage{authblk}
\usepackage[authoryear]{natbib}
\usepackage[right=2.5cm,left=2.5cm,top=2cm,bottom=3cm]{geometry}
\usepackage[font={small,it}]{caption} % the options makes the font italic in floats
\usepackage[colorlinks,linkcolor=blue,citecolor=blue,urlcolor=blue]{hyperref}
\usepackage{subfig}
\usepackage{mathtools}

\usepackage{multirow}
\usepackage{makecell} 
\usepackage{afterpage}
\RequirePackage[noend]{algpseudocode}
\RequirePackage{algorithm}
\algnewcommand{\algorithmicand}{\textbf{ and }} %\And command 
\algnewcommand{\And}{\algorithmicand} 

\title{ \textsc{An adaptive functional regression framework for spatially heterogeneous signals in spectroscopy} }
\author[1]{Federico Ferraccioli}
\author[2]{Alessandro Casa}
\author[3]{Marco Stefanucci}
\affil[1]{Department of Statistical Sciences, University of Padova, Italy}
\affil[2]{Faculty of Economics and Management, Free University of Bozen-Bolzano, Italy}
\affil[3]{Department of Economics and Finance, University of Rome Tor Vergata, Italy}
\date{}                     %% if you don't need date to appear

\begin{document}
\maketitle

\begin{abstract}
The attention towards food products characteristics, such as nutritional properties and traceability, as well as towards the adherence of production systems to environmental and ethical procedures, has risen substantially in the recent years. Consequently, we are witnessing an increased demand for the development of modern tools to monitor, analyse and assess food quality, security, and authenticity. Within this framework, an essential set of data collection techniques is provided by vibrational spectroscopy. In fact, methods such as Fourier near-infrared (NIR) and mid-infrared (MIR) spectroscopy have been often exploited to analyze different foodstuffs. Nonetheless, existing statistical methods often struggle to deal with the challenges presented by spectral data, such as their high-dimensionality, paired with strong relationships among the wavelengths. Therefore, the definition of proper statistical procedures accounting for the intrinsic peculiarities of spectroscopy data is paramount.

In this work, motivated by two dairy science applications, we propose an adaptive functional regression framework for spectroscopy data. The method stems from the trend filtering literature, allowing the definition of a highly flexible and adaptive estimator able to handle different degrees of smoothness. We provide a fast optimization procedure that is suitable for both Gaussian and non-Gaussian scalar responses, and allows for the inclusion of scalar covariates. Moreover, we develop inferential procedures for both the functional and the scalar component thus enhancing not only the interpretability of the results, but also their usability in real world scenarios. The method is applied to two sets of MIR spectroscopy data, providing excellent results when predicting both milk chemical composition and cows' dietary treatments. Moreover, the developed inferential routine provides relevant insights, potentially paving the way for a richer interpretation and a better understanding of the impact of specific wavelengths on milk features. 
\end{abstract}


\section{Introduction}
\label{sec:introduction}

In the past decades, increased consumers' attention towards food quality and security has fostered the development of new technologies to analyze different foodstuffs. More specifically, adherence to environmental-friendly procedures, product traceability, and quantification of the nutritional properties have become central topics on the agenda both of the public opinion and of the scientific community. Moreover, methods to assess food authenticity are increasingly important, as expensive products are often subject to fraud and adulteration. 

In this framework, commonly adopted methodologies often require lengthy and expensive laboratory extraction routines to collect data, thus jeopardizing their usefulness. As a consequence, other alternatives have recently been proposed to overcome these drawbacks, with vibrational spectroscopy techniques currently playing a pivotal role. Methods such as Fourier transform near-infrared (NIR) and mid-infrared (MIR) spectroscopy are known to be fast, relatively cheap, and nondisruptive ways to collect huge amounts of data on a plethora of materials. In fact, they have been used in different fields, ranging from medicine \citep{petrich2001mid,talari2017advances} and astronomy \citep{keller2006infrared,tennyson2019astronomical}, to food and animal science \citep{reid2006recent,berzaghi2009near,porep2015line}.

In this work, we focus specifically on MIR spectroscopy, where the light is passed through a sample of a given material at a sequence of wavelengths in the mid-infrared region, activating the sample's chemical bonds. This leads to an absorption of the energy from the light, the amount of which, evaluated at different wavelengths, creates the spectrum of the analyzed sample. Each spectrum contains an invaluable amount of information about the sample since, according to \emph{Beer-Lambert Law} \citep{beer1852bestimmung}, the absorption of energy is specific to atoms and molecules and is proportional to the concentration of the corresponding chemical substance. As a consequence, nowadays spectral data are being used to predict different characteristics of a given material. With specific reference to the dairy framework, the one considered in this work, MIR spectroscopy showed promising results in predicting traits such as milk coagulation properties \citep{visentin2016predictive}, fatty acids \citep{soyeurt2006estimating}, protein and lactose concentration \citep{de2014invited}, energy efficiency and intake \citep{mcparland2016potential}, as well as in discriminating between different cows' dietary treatments \citep{frizzarin2021application}.

Despite being widely used, the peculiar characteristics of spectroscopy data introduce statistical challenges that need to be addressed. First, spectral data lie in high-dimensional spaces, as each single spectrum usually consists of more than 1000 absorbance values measured at different wavelengths. Moreover, the relationships among variables are rather complex, often preventing the use of standard models developed for time-dependent data. In fact, even if adjacent spectral regions tend to be highly correlated, strong correlations are also observed among distant wavelengths, since the same chemical components can have several absorption peaks in distinct spectral regions. Lastly, as pointed out by \citet{politsch2020trend}, the underlying signal is often spatially heterogeneous. Therefore, flat spectral regions are often followed by more irregular ones, characterized by multiple peaks, posing cumbersome issues in the modelling process.

Both with regression and classification aims in mind, these data have been often analyzed by means of latent variable models. Methods such as \emph{Partial Least Squares} (PLS) and \emph{Principal Component Analysis} (PCA) have been widely used to tackle some of the mentioned problems. With a similar underlying rationale, \emph{Factor Analysis} has also been considered \citep[see e.g.][for a recent work]{casa2021parsimonious} since, allowing one to reduce the dimensionality of the data while focusing on a proper reconstruction of the correlation matrix, it seems particularly suitable for the framework. Recently, statistical and machine learning techniques have also been explored in order to relate spectral information to different milk traits \citep[see e.g.][]{FRIZZARIN20217438}. 

All these methods do not account for the peculiar structure of the spectral data and for the natural ordering among the variables, which can be considered by resorting to approaches pertaining to the functional data analysis setting \citep[FDA;][]{FDA1}. In fact, even if \citet{alsberg1993representation} suggested that spectra should be represented as continuous functions of wavelengths, in this framework functional approaches have been to some extent overlooked until relatively recently 
 \citep{saeys2008potential}. Some works that it is worth mentioning, even if not necessarily focused on MIR spectral data, are the ones by \citet{reiss2007functional,morris2008bayesian,zhao2012wavelet,yang2016smoothing,codazzi2022gaussian}. 

As briefly mentioned before, the varying degrees of smoothness of MIR spectroscopy data over their domain pose some challenges that need to be tackled when evaluating FDA strategies. \citet{saeys2008potential} suggest to adopt a basis approach with unequally spaced knots, with knot placement driven by subject-matter knowledge on the properties of the analyzed material. In this work, we take a different approach by considering \emph{trend filtering} as a building block of our approach \citep[see][for a discussion in a similar framework]{politsch2020trend}. 

\subsection{Trend Filtering}

Trend filtering is a signal reconstruction technique initially developed by \cite{kim_TF} and further studied, among others, by \cite{trend_filtering}. In the context of nonparametric regression, where data ${\bm{y} } = (y_1, y_2, \ldots, y_p )^{\top} \in \mathbb{R}^p$ are supposed to be generated by the model $y_i = f_0(\omega_i) + \varepsilon_i$, $i = 1, \dots, p$, trend filtering solves the following empirical problem 
\begin{equation}
\label{eq:problem0}
   \widehat{\bm{f}} = \arg\min_{\bm{f} \in \mathbb{R}^p} \lVert \bm{y} - \bm{f}\rVert_2^2 + \lambda \lVert \bm {D}^{(k+1)} \bm{f}  \rVert_1 
\end{equation}
where %${\boldsymbol \omega} = (\omega_1, \ldots, \omega_p), 
$\bm{f} = (f(\omega_1), \dots, f(\omega_p))^{\top},$ $\bm {D}^{(k+1)} \in \mathbb{R}^{(p-k-1)\times p}$ is the discrete difference matrix of order $k+1$, and $\lambda$ $>0$ is a tuning parameter. The resulting discretely estimated function $\widehat{\bm{f}}$ has a number of interesting properties, the most important being its adaptivity to local features of the true underlying function $f_0$. More precisely, the specification of the penalty yields a solution which, even if generally not sparse, exhibits a sparse $(k+1)$-th derivative. This behaviour resembles a spline function of degree $k$ which possesses continuous derivatives up to order $k-1$, while the $k$-th derivative is zero except for the points where polynomials meet, also known as \textit{knots} of such spline. As shown in \cite{trend_filtering}, for $k=1, 2$ the resulting estimated function is indeed a spline, while for $k>2$ it is \textit{close}, but not exactly equal, to a spline function with unequally spaced knots. The method is quite general thanks to the choice of $k$, e.g. with $k=0$ one obtains a stepwise function with a first derivative that is different from zero only where jumps lie. In fact, given the form of $\bm {D}^{(1)}$, in this specific case the penalty becomes $\sum_{j=1}^{p-1} |f(x_j) - f(x_{j+1})|$ and the problem is equivalent to the Fused Lasso \citep{fused_lasso}. With $k=1$ the second derivative is penalized, thus yielding an estimate that is piecewise linear. These and higher-order examples can be found in the original paper of \cite{trend_filtering}. A prominent instance is cubic trend filtering ($k=3$) that allows to fit to the data something very similar to a cubic spline with unequally spaced knots. The further relevance of this approach can be appreciated also from another point of view; in the literature, several adaptive estimation procedures have been proposed, mainly focusing on finding good sets of spline knots \citep[see e.g.,][]{rjmcmc_genovese,algo_knots}. The trend filtering approach implicitly overcomes this problem since, by solving the minimization in \eqref{eq:problem0} for a given $\lambda$, only a number $p_{\lambda} < p$ of knots are selected; the entire path spans a range of nested solutions without the need of a forward/backward knot search algorithm.

%Starting from Trend Filtering we extend the main ideas to functional regression with Gaussian scalar response, by developing an estimator able to infer spatially inhomogeneous regression functions. We further investigate this modeling strategy in the case of a \textit{partial} functional linear model, i.e. by including a set of scalar covariates. Finally, we propose a last extension useful when the response variable is non-gaussian, as for example presence/absence or count data.

In this paper, after a brief description of the analyzed data in Section \ref{sec:data description}, in Section \ref{sec:method} we extend the main  trend filtering concepts to functional regression with Gaussian scalar response, developing an estimator able to infer spatially inhomogeneous regression functions. We further investigate this modeling strategy in the case of a \textit{partial} functional linear model, i.e. by including a set of scalar covariates and propose an extension intended for non-Gaussian response, as for example presence/absence or count data.
Efficient estimation algorithms are presented in Section \ref{sec:comp}, followed by a simulation study in Section \ref{sec:simstudy}. In Section \ref{sec:application} we present the result of the analysis on real data, while in Section \ref{sec:conclusions} we draw some final conclusions and remarks. 

\section{Mid-infrared spectroscopy data}
\label{sec:data description}

\begin{figure}
    \centering
    \includegraphics[width = \linewidth]{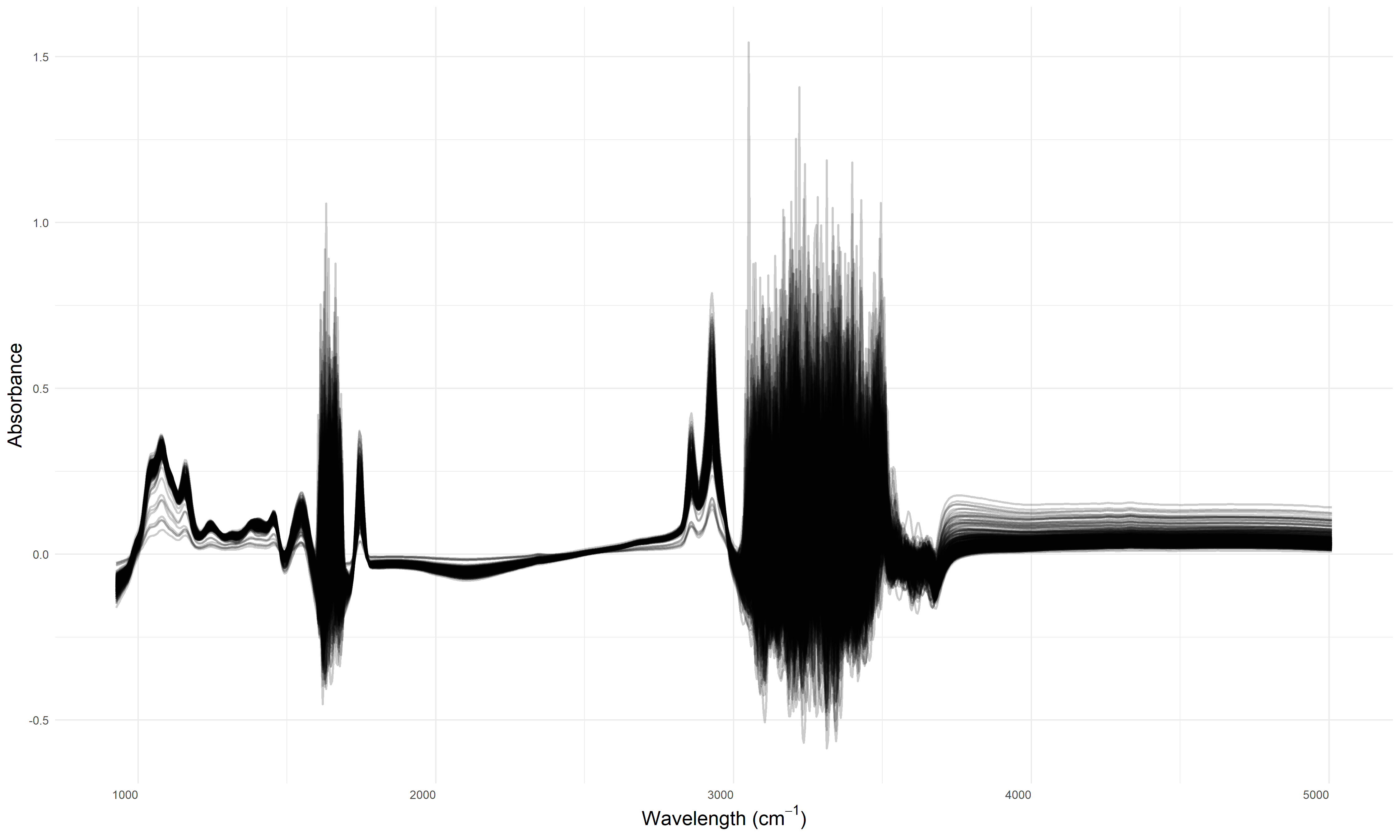}
    \caption{Mid-infrared spectra in the region from $925 cm^{-1}$ to $5010 cm^{-1}$, corresponding to the first case study.}
    \label{fig:dat}
\end{figure}

In this study, we consider two different sets of mid-infrared spectroscopy data.

The first data set consists of a collection of 730 milk samples produced from 622 cows from different research herds in Ireland between August 2013 and August 2014. All the animals involved in the study were following a predominantly grass-based diet, with relevant heterogeneity in terms of number of parities and stage of lactation. Samples were collected during morning and evening milking and analyzed using a MilkoScan FT6000 (Foss Electronic A/S, Hillerød, Denmark). The resulting spectra consists of 1060 absorbance observations in the mid-infrared light region (see Figure \ref{fig:dat}). Furthermore, some additional information is available such as the date and the time (am/pm) of the milkings, the breed, the number of parities and the days in milk for all the cows involved in the study. Note that some milk related traits have been collected by means of wet chemistry techniques. Among these traits are included both technological, such as rennet coagulation time and heat stability, and protein related ones, as for example $\kappa$-casein and $\alpha_{S1}$-casein. In the analyses reported in Section \ref{sec:application}, we focus on the prediction of the $\kappa$-casein. Lastly, we retain only one observation per cow, thus working with 622 milk spectra. For a more complete description of the data collection process and of the data themselves, readers can refer to \citet{visentin2015prediction}. 

On the other hand, the second data set considered has been collected at the Teagasc Moorepark Dairy Research Farm (Fermoy, Co.Cork, Ireland), in an experiment designed by \citet{o2016effect}, which represents the first study of its kind in Ireland and, to the best of our knowledge, in the world. Further information on the experimental setting can be found in \citet{o2016quality,o2017effect}. The data consist of MIR spectra, comprising 1060 wavelengths in the region from 925cm$^{-1}$ and 5010cm$^{-1}$, obtained by analyzing 4320 milk samples using a Pro-Foss FT6000 series instrument. A total number of 120 Holstein-Friesian cows have been involved in the study, and milked twiced daily in the morning and in the afternoon in three consecutive years (2015, 2016 and 2017). The data collection scheme has been carried out in a balanced way, both in terms of the year and of the number of cattle' parities. Moreover, we restrict our attention to the samples collected from May to August, since in the summer period there is the highest prevalence of grass growth. For each of the years considered, the cattle were randomly assigned to a specific dietary treatment for the entire lactation period. The treatment diets included grass (GRS), with cows mantained outdoors on a perennial ryegrass sward only, clover (CLV), consisting of perennial ryegrass with 20\% white clover sward only, and total mixed ration (TMR), where cows were mantained indoors with nutrients combined in a single mix consisting of grass and maize silage and concentrates. In this work, given the strong compositional similarities between GRS and CLV diets, these two classes have been merged to create a general pasture-based diet group. As a consequence the final data set consists of 2931 samples from pasture-fed cattles and 1389 from TMR-fed ones. Lastly, some additional information on fat, protein, and lactose content has been obtained by calibrating the FT6000 against wet chemistry results and is available for the milk samples considered.

\section{Proposed methodology}
\label{sec:method}

Given the data $\mathcal{D} = \{ X_i(\omega), y_i \}_{i=1}^n$, we assume that $y_1, \ldots, y_n$ are scalar values drawn from a Gaussian random variable $y_i|X_i(\omega) \sim N(\mu_i, \sigma^2)$ and $X_1(\omega), \dots, X_n(\omega)$ are realizations of a functional covariate. We model the conditional expected value of $y_i$ as $\mathbb{E}(y_i|X_i(\omega)) = \mu_i = \int X_i(\omega) f(\omega) d\omega$ where $f(\omega)$ is an unknown regression function. This leads to the functional linear model 
\begin{equation*}
%\label{eq:model1}
 y_i = \int X_i(\omega) f(\omega) d\omega + \varepsilon_i\, ,  
\end{equation*}
with $\varepsilon_i \sim N(0, \sigma^2)$ being an additive noise term. There are several works in the functional data analysis literature devoted to estimation of this model, ranging from basis approaches \citep{FDA1}, penalized stategies \citep{funreg_splines} and functional principal component representations \citep{funreg_fpca}. A complete review of different estimation methodologies is outside the scope of this work, and readers may refer to \cite{review_morris} for a general overview of such methods; however, here we spend some words on the penalization approach, which shares some features with our proposal.  

Smoothing splines are the most popular tool in the family of penalized estimators. In the context of functional regression, given the data $\mathcal{D}$, they are obtained as the solution of the following optimization problem:
\begin{equation}
\label{eq:problem_splines}
\widehat{\bm{f}} = \arg\min_{\bm{f} \in \mathbb{R}^p} \lVert \bm{y} - \boldsymbol{X}\bm{f}\rVert_2^2 + \lambda \lVert \bm {D}^{(2)} \bm{f} \rVert_2^2 \,, 
\end{equation}
where ${\bm y} = (y_1, \ldots, y_n)^{\top}$ is the vector of scalar responses, $\boldsymbol{X} = (X_1({\bm \omega}), \ldots, X_n({\bm \omega}))^{\top}$ is the matrix of functional data observed on a regular grid ${\bm \omega}= (\omega_1, \ldots, \omega_p)$ and $\bm{D}^{(2)}$ is the matrix of second order discrete differences. The approach is justified as a discrete counterpart of a certain variational problem and provides as a solution a natural cubic spline with knots at observation points, see \cite{wahba_book}, \cite{funreg_splines} and \cite{goldsmith2011penalized} for details. The amount of penalization, managed by the tuning parameter $\lambda$, represents a trade-off between two extreme solutions, one being a completely wiggly interpolating function ($\lambda=0$) and the other being a constant ($\lambda = \infty$). Despite being very intuitive and easy to implement, this estimator lacks the local adaptivity property, i.e. the resulting estimated curve is not able to capture the different levels of smoothness of the true function $f$. Therefore, we propose to rely on a different penalization strategy and estimate the regression curve by
\begin{equation}
\label{eq:problem1}
\widehat{\bm{f}} = \arg\min_{\bm{f} \in \mathbb{R}^p} \lVert \bm{y} - \boldsymbol{X}\bm{f}\rVert_2^2 + \lambda \lVert \bm{D}^{(k+1)} \bm{f}  \rVert_1  .
\end{equation}
This expression represents a generalization of the trend filtering loss function, with the design matrix no longer being the $p \times p$ identity matrix \citep{trend_filtering}, but an $n \times p$ matrix of discretely observed functional data. 
The key difference between optimization problems (\ref{eq:problem_splines}) and (\ref{eq:problem1}) is that in the latter, thanks to the $\ell_1$-penalty applied on certain discrete derivative of $f$, we are able to estimate the regression function taking care of its local features. Indeed, similarly to the original trend filtering estimate, the function that minimizes (\ref{eq:problem1}) is equal (when $k=1, 2$) or very similar (when $k > 2$) to a spline function with unequally spaced knots along its domain. Clearly, the smoothing splines approach, not being able to adapt to local features of the curve, either misses its smooth or its wiggly parts, depending on the selected degrees of freedom. This is a consequence of the  regularization term which does not produce spatial heterogeneity nor knot selection. This makes the approach in (\ref{eq:problem1}) particularly appealing for spectroscopy data, where the effect of a functional covariate (e.g. the MIR spectrum) on a scalar variable (e.g. a material trait) can be studied with particular attention to local effects.

\subsection{Extensions}\label{sec:extensions}

In the current framework, when scalar covariates are available alongside the functional one, their inclusion in the modelling strategy can bring additional information useful to predict the response variable. More formally, in this setting we denote the observed data by 
$\mathcal{D} = \{ X_i(\omega), y_i,\bm {z}_i  \}_{i=1}^n$, 
%$\mathcal{D} = \{ X_i(\omega), y_i, z_{1i}, \ldots, z_{ri} \}_{i=1}^n$, 
with $\bm {z}_i = (z_{i1}, \ldots, z_{ir})^{\top}$ being a set of $r$ scalar covariates corresponding to the $i$-th observation. We then model the conditional expected value of $y_i$ as $\mathbb{E}(y_i|X_i(\omega), {\bm z}_i) = \mu_i = \int X_i(\omega) f(\omega) d\omega + \sum_{j=1}^r z_{ij} \gamma_j $, where $\{ \gamma_j \}_{j=1}^r$ are unknown regression coefficients.  For such data structure, it is worth considering the partial functional linear model \citep{partial_flm}
\begin{equation*}
    y_i = \int X_i(\omega) f(\omega) d\omega + \sum_{j=1}^r z_{ij} \gamma_j +  \varepsilon_i \, .
\end{equation*}  
Note that, if needed, the intercept can be easily included in the model as a constant scalar covariate. Following the trend filtering paradigm, we propose to estimate the function $f$ and the vector $\bm{ \gamma} = (\gamma_1, \ldots, \gamma_r)^{\top}$ by solving the optimization problem 
\begin{equation}
\label{eq:problem2}
\widehat{\bm{\theta}} = \arg\min_{\bm{\theta} \in \mathbb{R}^{p+r}} \lVert \bm{y} - \Tilde{\boldsymbol{X}}\bm{\theta} \rVert_2^2 + \lambda \lVert \Tilde{\bm D}^{(k+1)} \bm{\theta}  \rVert_1  \, ,
\end{equation}
where $\Tilde{\boldsymbol{X}} = [ \boldsymbol{X} | \boldsymbol{Z}] \in \mathbb{R}^{n \times (p + r)}$, $\boldsymbol{Z} = (\boldsymbol{z}_1, \dots, \boldsymbol{z}_n)^{\top} \in \mathbb{R}^{n \times r}$, $\bm{\theta} = ({\bm f}^{\top}, {\bm \gamma}^{\top})^{\top} \in \mathbb{R}^{p+r}$ and $\Tilde{\bm D}^{(k+1)} = [{\bm D}^{(k+1)} | \bm{0}_{(p-k-1)\times r}] \in \mathbb{R}^{(p-k-1)\times(p+r)}$. Note that, with this formulation, the penalty does not affect the parametric part of the model. When $r$ is large, one can include an $\ell_1$-penalty for the vector $\bm{\gamma}$ in order to achieve sparsity in the estimated coefficients; see, for instance, \cite{kong2016partial}. Since the application presented in Section \ref{sec:application} involves a small set of covariates, this potential extension has not been pursued in this work. 

An additional generalization of our proposal is required when assumption $y_i \sim N(\mu_i, \sigma^2)$ is not met because the scalar responses $y_1, \ldots, y_n$ are generated by some other distribution. For example, for count data, we can assume $y_i \sim Poisson(\lambda_i)$, and for presence/absence data $y_i \sim Bernoulli(\pi_i)$. In these settings, where a functional linear model is not adequate, a generalized functional linear model \citep{james2002glm, gen_flm, goldsmith2011penalized} can be applied. In particular, we assume that $g(\mathbb{E}(y_i|X_i(\omega))) = g(\mu_i) = \int X_i(\omega)f(\omega) d\omega$, with $g(\cdot)$ being a suitably chosen link function. Now the empirical minimization problem is recasted as 
\begin{equation}
\label{eq:problem3} 
\widehat{\bm{f}} = \arg\min_{\bm{f} \in \mathbb{R}^p} L({\bm y}; {\bm X}{\bm f}) + \lambda \lVert \bm{D}^{(k+1)} \bm{f}  \rVert_1  ,
\end{equation}
where the loss function $L(\bm{y}; \bm{X}\bm{f})$ depends upon the distribution of the response variable. The objective is now represented by a nonlinear function of the unknown parameter $\bm{f}$ and its direct minimization is usually not straightforward. As a consequence, in Section \ref{sec:comp} we  present a clever modification of the proposed algorithm to deal with the modified loss appearing in (\ref{eq:problem3}). 
%construct an iterative algorithm based on reweighted least squares. 
Lastly note that in the presence of explanatory scalar covariates and a non-Gaussian response, the last two specifications can be combined together by adjusting (\ref{eq:problem3}) as it has been done in equation (\ref{eq:problem2}) for problem (\ref{eq:problem1}).

Another potential extension would be to combine two (or more) penalties in the optimization problem. This allows to estimate functions that exhibit a complex behaviour, typically piecewise polynomials of different order. The loss function in this context is 
\begin{equation}
\label{eq:problem4}
\widehat{\bm{f}} = \arg\min_{\bm{f} \in \mathbb{R}^p} \lVert \bm{y} - \boldsymbol{X}\bm{f}\rVert_2^2 + \lambda_1 \lVert \bm{D}^{(k+1)} \bm{f}  \rVert_1 + \lambda_2 \lVert \bm{D}^{(\ell+1)} \bm{f}  \rVert_1 ,
\end{equation}
where $k$ and $\ell$ are integers and $\lambda_1$, $\lambda_2$ are regularization parameters. This modification can be employed when additional scalar covariates are observed and/or when the distribution of the response is not Gaussian. In section \ref{sec:comp} we will illustrate how to solve problem (\ref{eq:problem4}) with the same toolbox used for the other cases.

%%%%%%%%%%%%%%%%%%%%%%%%%%%%%%%%%%%%%%%%%%%%%%%%%%%

\subsection{Inference}\label{sec:Inference}
In this section, we describe a strategy to build confidence intervals for most of the pointwise estimates introduced in the previous section. Given the complexity of functional regression models, inferential procedures have sometimes been overlooked, with the focus often being on pointwise estimation. Nonetheless, the introduced procedure represents a key component that improves the usability of the methodology in real world scenarios. 
%This is indeed a key component that improve the usability of the proposed methodology in real world scenarios. Inferential procedures usually do not represent the main target in functional regression models, focusing more on pointwise estimation. This is also due to the complexity of such functional models. 
In the case of the trend filtering framework, the construction of confidence intervals and confidence bands can be addressed via bootstrap procedures. Standard frequentist inference is not suitable, since the distribution of the trend filtering estimator is non-Gaussian, even when the observational noise is Gaussian \citep{politsch2020trend}. Here we propose a Wild bootstrap procedure \citep{mammen1993bootstrap}, that is particularly appropriate in high-dimensional regression models when the noise distribution is unknown. Briefly, the idea behind the Wild bootstrap is to construct an auxiliary random variable with zero mean and unit variance (and ideally higher moments equal to 1). This random variable is then used to define a transformation of the observed residuals that gives a valid bootstrap sample (see Algorithm \ref{alg:wild}). %{\color{red} qui valutiamo se tenere schemino o spiegare i passaggi nel testo, il che renderebbe forse anche la sezione leggermente più lunga.}

A classical choice for the auxiliary random variable is the two point distribution suggested in \cite{mammen1993bootstrap}, that is
\begin{equation}\label{eq:auxiliary}
    u_{i}^{*} = 
            \begin{cases}
                \hat{\epsilon}_{i}(1 + \sqrt{5})/2 \quad \text{with probability} (1 + \sqrt{5})/(2\sqrt{5}),\\
                \hat{\epsilon}_{i}(1 - \sqrt{5})/2 \quad \text{with probability} (\sqrt{5} - 1)/(2\sqrt{5}).
            \end{cases}
\end{equation}
Other examples are the Rademacher distribution, that takes values ${1, -1}$ with equal probability, the Uniform distribution on the interval $[-\sqrt{3}, \sqrt{3}]$, or various transformations of the Gaussian distribution \citep{mammen1993bootstrap}. In general, since it is not possible to define a random variable that has mean 0, variance 1, and all higher moments equal to 1, the different choices lead to different values for the third and the fourth moment. For instance, the third and fourth moments of the Rademacher distribution are 0 and 1, respectively, while the two point distribution defined above have third and fourth moments 1 and 2, respectively. The specific choice is generally driven by considerations on the symmetry of the observed residuals.

\begin{algorithm}[t]
\caption{Wild bootstrap} \label{alg:wild}
\begin{algorithmic}[1]
    \Require $\bm{X}, \bm{y}, \bm{D}, \lambda$
    \State Compute the trend filtering estimate at the observed data points
    \State Let $\widehat{\epsilon}_{i} = y_{i} - \widehat{y}_{i} = y_{i} - \sum_{j = 1}^{p}X_{i}(\omega_j)\widehat{f}(\omega_j)$
    \For{b \textbf{in} 1:B}
    \For{\textbf{all} i}
        \State Define a bootstrap sample by sampling from the following distribution
        \begin{equation*}
            y_{i}^{*} = \widehat{y}_i + u_{i}^{*} \quad i = 1, \dots, n,
        \end{equation*}
            $\quad \quad \;\;\;\;$ with $u_{i}^{*}$ defined as in \eqref{eq:auxiliary}.
    \EndFor
    \State Let $\widehat{\bm{f}}^{(b)} = (\widehat{f}^{(b)}(\omega_1), \dots, \widehat{f}^{(b)}(\omega_p))$ be the estimate from the bootstrap sample.
    \EndFor
    \State Compute the bootstrap-based confidence bands as in \eqref{eq:wb_int}.
\end{algorithmic}
\end{algorithm}
Given the full bootstrap estimate set $\{\widehat{\bm{f}}^{(b)}\}_{b=1}^B$ , for any $\alpha \in (0, 1)$, we can define a $(1 - \alpha)$ quantile-based pointwise variability band as
\begin{equation}
    \label{eq:wb_int}
    V_{1 - \alpha}(f(\omega_j)) = \left(\widehat{f}_{\alpha/2}(\omega_j), \widehat{f}_{1 - \alpha/2}(\omega_j)\right), 
\end{equation}
%\ff{qua non sono sicuro, forse meglio specificare l'intervallo puntale per $V_{1 - \alpha}(\omega)$ e definirlo come $\left(\sum_{j = 1}^{q}\hat{\beta}_{j}^{\alpha/2} \phi_{j}(\omega), \sum_{j = 1}^{q}\hat{\beta}_{j}^{1 - \alpha/2} \phi_{j}(\omega) \right)$, così l'intervallo è definito per qualsiasi valore.} \\
where 
\begin{equation*}
    \widehat{f}_{\gamma}(\omega_j) = \underset{g}{\inf}\left\{g: \frac{1}{B}\sum_{b = 1}^{B}\mathbb{I}(\widehat{f}^{(b)}(\omega_j) \leq g) \geq \gamma \right\}, \quad \text{for all} \quad j = 1, \dots, p.
\end{equation*}

%%%%%%%%%%%%%%%%%%%%%%%%%%%%%%%%%%%%%%%%%%%%%%%%%%%

\section{Optimization procedure}
\label{sec:comp}

In the literature, several algorithms for solving the original trend filtering problem have been proposed; see, among others, \cite{kim_TF}, \cite{trend_filtering} and \cite{gen_lasso}. Some of these algorithms are not directly generalizable to our context, where the presence of the $n \times p$ data matrix $\bm X$ makes the optimization task more challenging.
%there is a $n \times p$ full matrix $\bm X$ in place of the identity matrix.
To solve problem (\ref{eq:problem1}), we rely on the Alternating Direction Method of Multipliers (ADMM) framework and consider an extension of the approach by \cite{ramdasADMM} where a specialized acceleration scheme is proposed. 

ADMM algorithms are a wide class of algorithms particularly useful for solving constrained problems of the form
\begin{align}
 \text{minimize} & \quad f({\bm \alpha}) + g({\bm \delta}) \label{eq:admm1} \, ,\\
 \text{subject to} & \quad {\bm A} {\bm \alpha} + {\bm B} {\bm \delta} + {\bm c} = 0 \nonumber.
\end{align}
A general ADMM algorithm proceeds by minimizing the augmented Lagrangian of (\ref{eq:admm1}). Since the objective function is separable, minimization can take place in an alternate fashion. ADMM approaches are largely used in penalized estimation schemes, which can often be recasted as in (\ref{eq:admm1}), leading to a faster optimization thanks to variable splitting. Specifically, the problem in (\ref{eq:problem1}) can be stated as
\begin{align}
 \text{minimize} & \quad \lVert {\bm y} - {\bm X}{\bm \alpha} \rVert_2^2  + \lVert {\bm \delta}\rVert_1 \label{eq:admm2} \, ,\\
 \text{subject to} & \quad {\bm D}^{(k+1)} {\bm \alpha} - {\bm \delta} = 0 \nonumber.
\end{align}
where $f({\bm \alpha}) = \lVert {\bm y} - {\bm X}{\bm \alpha} \rVert_2^2  $ is the $\ell_2$-loss and $g({\bm \delta}) = \lVert {\bm \delta}\rVert_1$ the $\ell_1$-norm. As shown in \cite{boydADMM}, updates for the parameters are straightforward. In fact, since $f$ is quadratic, the update step for ${\bm \alpha}$ has a least squares form and the ${\bm \delta}$ update amounts in soft-thresholding a given vector. 
%Explicit formulas for the generic $t+1$ step are the following: 
%
%\begin{align*}
%{\bm \alpha}^{t+1} & = ({\bm X}^{\top}{\bm X} + \rho ({\bm D}^{(k+1)})^{\top}{\bm D}^{(k+1)})^{-1}({\bm X}^{\top} {\bm y} + \rho ({\bm D}^{(k+1)})^{\top} ({\bm \delta}^{t} - {\bm u}^{t})) \\
% {\bm \delta}^{t+1} & = \mathcal{S}_{\lambda/\rho}({\bm D}^{(k+1)}{\bm \alpha}^{t+1} + {\bm u}^{t})\\
% {\bm u}^{t+1} & = {\bm u}^{t} + {\bm D}^{(k+1)}{\bm \alpha}^{t+1} - {\bm \delta}^{t+1}
%\end{align*}
%
%where ${\bm D}^{(k+1)}$ is the $(k+1)$th discrete derivative matrix, $\mathcal{S}_{\kappa}(x) = (1 - \kappa/|x|)_+ x$ is the soft-thresholding operator and $\rho$ is the penalty parameter of the augmented Lagrangian. 
Although these updating rules could be applied, an acceleration scheme for such problem is exploited. In fact, as demonstrated in \cite{ramdasADMM}, a different parametrization of the ADMM can save computational time due to the existence of efficient algorithms for the constant-order trend filtering problem. The idea is to reformulate the problem as follows. 
\begin{align}
 \text{minimize} & \quad \lVert {\bm y} - {\bm X}{\bm \alpha} \rVert_2^2  + \lVert{\bm D^{(1)}}{\bm \delta} \rVert_1  \, ,\label{eq:admm3}\\
 \text{subject to} & \quad {\bm D}^{(k)} {\bm \alpha} - {\bm \delta} = 0 \nonumber.
\end{align}
where $\bm D^{(1)}$ is the discrete difference matrix of order 1. The reader can verify the equivalence between problem (\ref{eq:admm2}) and problem (\ref{eq:admm3}). Hereafter, we derive the specialized parameter updates needed for the generic $t+1$ iteration: 
\begin{align}
{\bm \alpha}^{t+1} & = ({\bm X}^{\top}{\bm X} + \rho ({\bm D}^{(k)})^{\top}{\bm D}^{(k)})^{-1}({\bm X}^{\top} {\bm y} + \rho ({\bm D}^{(k)})^{\top} ({\bm \delta}^{t} - {\bm u}^{t}))  \label{eq:RAM1}\, ,\\
 {\bm \delta}^{t+1} & = \arg\min \lVert {\bm D}^{(k)}{\bm \alpha}^{t+1} + {\bm u}^{t} - {\bm \delta}^{t}  \rVert_2^2 + \lambda/\rho \lVert {\bm D^{(1)}} {\bm \delta}^{t}   \rVert_1  \label{eq:RAM2}\, ,\\
 {\bm u}^{t+1} & = {\bm u}^{t} + {\bm D}^{(k)}{\bm \alpha}^{t+1} - {\bm \delta}^{t+1} \label{eq:RAM3}\, .
\end{align}
The update for ${\bm \delta}$ is much more involved than a simple soft-thresholding and requires solving a new constant-order trend filtering problem, that is, a one-dimensional fused lasso problem. However, fast solutions are available by employing the dynamic programming solver by \cite{DP_johnson} or the proposal by \cite{davies_kovac_algo} based on the taut string principle. \cite{ramdasADMM} showed the superiority of this specialized ADMM formulation over the classical one in terms of convergence rates: the single operation is more expensive than the one in the usual parametrization, but convergence is achieved in fewer iterations, leading to an overall gain in terms of computational time.

The parameter $\rho$ is sometimes made adaptive by allowing a different value at each iteration to speed up the learning process. Using a varying $\rho$, one has to compute $({\bm X}^{\top}{\bm X} + \rho^t {\bm D}^{\top}{\bm D})^{-1}$ at each iteration of the algorithm, and this can be prohibitive even for moderate dimensions. With a fixed $\rho$ instead, one can precompute the quantity $({\bm X}^{\top}{\bm X} + \rho {\bm D}^{\top}{\bm D})^{-1}$ that is never modified by the updating rules at the expense of some more iterations. In our implementation, we found this last approach faster and, in particular, we followed \cite{ramdasADMM} setting $\rho = \lambda$, which led to stable solutions.
%The parameter $\rho$ is sometimes made adaptive by allowing a different value at each iteration to speed up the learning process. With a fixed $\rho$ value, one can pre-compute the quantity $({\bm X}^{\top}{\bm X} + \rho {\bm D}^{\top}{\bm D})^{-1}$, which is time consuming and never modified by the updating rules while, using a varying $\rho$, one should compute $({\bm X}^{\top}{\bm X} + \rho^t {\bm D}^{\top}{\bm D})^{-1}$ at each iteration. Depending on the dimension of the problem, the latter approach could be faster. In our implementation we found this choice less advantageous thus we fix the $\rho$ parameter, in particular the value $\rho = \lambda$ leads to stable solutions, as noted also by [REF, ramdas]. 
From a practical point of view, often the entire solution path is needed as a function of $\lambda$. In this case, a speed up is made by considering warm starts i.e., by starting the algorithm from the solution obtained for the previous value of the regularization parameter.

Lastly, note that slight modifications are needed in the presence of scalar covariates: the problem is stated as the minimization of $ \lVert {\bm y} - \Tilde{\bm X}{\bm \alpha} \rVert_2^2  + \lVert{\bm D^{(1)}}{\bm \delta} \rVert_1 $ subject to $ \Tilde{\bm D}^{(k)} {\bm \alpha} - {\bm \delta} = 0 $ and the updating rules are the same as (\eqref{eq:RAM1} -- \eqref{eq:RAM3}) except for the substitution of $ \Tilde{\bm X}$ and $ \Tilde{\bm D}^{(k)} $ in place of $ {\bm X}$ and $ {\bm D}^{(k)} $.
 
For the generalized functional linear model, we develop an iterative reweighted penalized least squares approach based on the alternation of a Newton step and an ADMM step. Specifically, problem (\ref{eq:problem3}) can be written as
\begin{align}
 \text{minimize} & \quad L({\bm y}; {\bm X}{\bm \alpha})  + \lVert{\bm D^{(1)}}{\bm \delta} \rVert_1 \, ,\\
 \text{subject to} & \quad {\bm D}^{(k)} {\bm \alpha} - {\bm \delta} = 0 \, .\nonumber
\end{align}
In the first step of the algorithm, given the current estimate ${\bm \alpha}^t$ we approximate the generic loss function $L({\bm y}; {\bm X}{\bm \alpha})$ around ${\bm \alpha}^t$ by a quadratic loss $\lVert \Tilde{\bm{y}}^t - \Tilde{\boldsymbol{X}}^t\bm{\alpha}\rVert_2^2$, where $ \Tilde{\bm y}^t = ({\bm W}^t)^{1/2} {\bm s}^t = ({\bm W}^t)^{1/2} ({\bm X}{\bm \alpha}^t + ({\bm V}^t)^{-1}({\bm y} - {\bm \mu}^t))$ and $ \Tilde{\bm X}^t = ({\bm W}^t)^{1/2} {\bm X}$, building a penalized least squares problem. This step is intended as a Fisher scoring update. 
%$$  L({\bm y}; {\bm X}{\bm \alpha}) + \lambda \lVert \bm{D}^{(k+1)} \bm{\alpha}  \rVert_1 \approx \lVert \Tilde{\bm{y}}^t - \Tilde{\boldsymbol{X}}^t\bm{\alpha}\rVert_2^2 + \lambda \lVert \bm{D}^{(k+1)} \bm{\alpha}  \rVert_1 $$
%= L(\Tilde{{\bm y}}^k, \Tilde{{\bm X}}^k)$$
%$$ \text{where} \quad \Tilde{\bm y}^t = ({\bm W}^t)^{1/2} {\bm s}^t = ({\bm W}^t)^{1/2} ({\bm X}{\bm \alpha}^t + ({\bm V}^t)^{-1}({\bm y} - {\bm \mu}^t))$$
%$$ \text{and} \quad \Tilde{\bm X}^t = ({\bm W}^t)^{1/2} {\bm X}$$
%where $ \Tilde{\bm y}^t = ({\bm W}^t)^{1/2} {\bm s}^t = ({\bm W}^t)^{1/2} ({\bm X}{\bm \alpha}^t + ({\bm V}^t)^{-1}({\bm y} - {\bm \mu}^t))$ and $ \Tilde{\bm X}^t = ({\bm W}^t)^{1/2} {\bm X}$. 
The quantities ${\bm \mu} = \mathbb{E}({\bm y}| {\bm X})$, ${\bm V} = V({\bm \mu})$ and ${\bm W} = V({\bm \mu})^{-1} (g'({\bm \mu}))^{-2}$ depend on the random variable characterizing the response. For example, if $y_i \sim Bernoulli(\pi_i)$ we have $\mu_i = \pi_i = \text{exp}\{\int X_i(\omega) f(\omega) d\omega\}/(1 + \text{exp}\{\int X_i(\omega) f(\omega)d\omega\})$, $V(\mu_i) = \mu_i(1 - \mu_i)$, $g'(\mu_i) = 1/V(\mu_i)$ and $W_{ii} = V(\mu_i)$ while if $y_i \sim Poisson(\lambda_i)$ we have $\mu_i = \lambda_i = \text{exp}\{\int X_i(\omega) f(\omega) d\omega\}$, $V(\mu_i) = \mu_i$, $g'(\mu_i) = 1/V(\mu_i)$ and $W_{ii} = V(\mu_i)$. 

In the second step we solve the penalized problem by applying ADMM updates (\eqref{eq:RAM1} -- \eqref{eq:RAM3}) until convergence, just by replacing ${\bm X}$ with $\Tilde{\bm X}^t$ and ${\bm y}$ with $\Tilde{\bm y}^t$, thus obtaining ${\bm \alpha}^{t+1}$. The two steps are repeated until some stopping criterion is achieved and the final estimator is obtained.

Lastly, note that it is possible to use the same machinery presented in this section for the multiple penalty approach too, by stacking the two matrices ${\bm D}^{(k+1)}$ and ${\bm D}^{(\ell+1)}$ to form $\Tilde{\bm D}$, which will replace the difference matrix in the ADMM updates.

%%%%%%%%%%%%%%%%%%%%%%%%%%%%%%%%%%%%%%%%%%%%%%%%%%

\section{Simulation study}
\label{sec:simstudy}

\begin{figure}
    \centering
    \includegraphics[width = 0.99\linewidth]{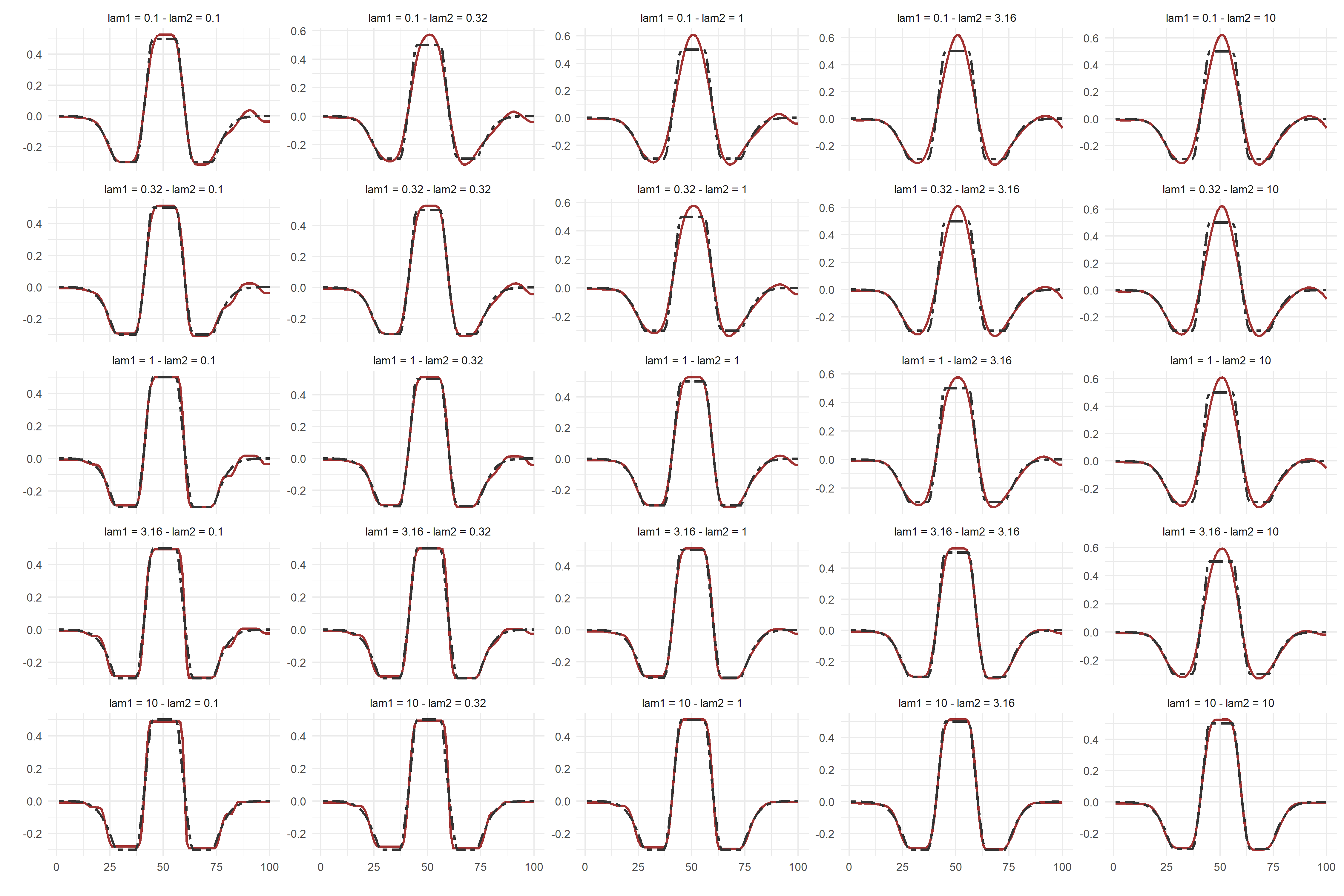}
    \caption{Estimated functional coefficient for $f_3(\omega)$, with different combinations of $(\lambda_1, \lambda_2)$ in the mixed penalty case $k = 0, l = 3$. The dashed-black lines correspond to the true function, while the solid-red lines correspond to estimates. From top to bottom, the estimated functional coefficient approaches a piecewise-constant function; from left to right, it approaches a cubic function.}
    \label{fig:truncated}
\end{figure}

In this section, we assess the performance of the proposed methods by means of simulations. We first generate a sample of functional data $\{X_i(\omega) \}_{i=1}^n$ from a B-spline basis with 10 equispaced internal knots, drawing each coefficient from a standard normal distribution. The resulting functions are evaluated on an equispaced grid of $p=100$ points in order to form the $n \times p$ matrix ${\bm X}$, and then kept fixed in all simulation repetitions. We define several scenarios that can be addressed with one of the methods previously described in the following way.

In scenario a), given the sample of functional data $\{X_i(\omega) \}_{i=1}^n$ we generate a sample of scalar responses from a Gaussian distribution $y_i \sim N(\mu_i, \sigma^2)$ where the expected value depends linearly on the functional covariate, i.e. $\mu_i = \int X_i(\omega) f(\omega) d\omega$. We set $n=250$ and a signal-to-noise ratio equal to 4. %For this scenario we estimate $f(\omega)$ solving problem \eqref{eq:problem1}.

%Scenario b) is the same as scenario a) with the exception that $n=50$. This setting completes the previous one by considering $p > n$. {\color{red} [dire altro? aggiungere norma L2?]}

In scenario b), in addition to the functional data sample, we also consider a set of scalar covariates $z_{1i}, \dots, z_{ri}$ for each observational unit. These covariates are generated from a standard Gaussian distribution and are independent of each other. Then, a sample of scalar responses is obtained from a Gaussian distribution $y_i \sim N(\mu_i, \sigma^2)$ where the expected value depends linearly on the functional and scalar covariates, i.e. $\mu_i = \int X_i(\omega) f(\omega) d\omega + \sum_{j=1}^r z_{ji} \gamma_j$. We set $n=250$, $r=5$, ${\bf \gamma} = (2, -1, 1, 0, 0)$ and a signal-to-noise ratio equal to 4. %For this scenario we estimate $f(\omega)$ and ${\bf \gamma}$ solving problem \eqref{eq:problem2}.

In scenario c), given the functional data sample, we generate scalar responses from a Bernoulli distribution $y_i \sim Bernoulli(\pi_i)$ where $g(\pi_i) = \text{logit} \{ \pi_i \}$  depends linearly on the functional covariate, i.e. $\text{logit} \{ {\pi_i}\} = \int X_i(\omega) f(\omega) d\omega$. We set $n=250$. %For this scenario we estimate $f(\omega)$ solving problem \eqref{eq:problem3}.

We combine each described scenario with three different specifications of the unknown regression function $f(\omega)$. In detail, $f_1(\omega)$ is a piecewise cubic function in $[0, 1]$ built from a cubic B-spline basis with 3 internal knots at $0.2, 0.75$ and $ 0.9$, $f_2(\omega)$ is the classical mexican hat function 
$$ f_2(\omega) = (1 - \omega^2)\text{exp}\{-\omega^2/2\} \quad \text{for } \omega \in [-5, 5], $$
and $f_3(\omega)$ is the same function with truncated peaks
$$ f_3(\omega) = \begin{cases}
        f_2(\omega) & \text{if } f_2(\omega) \in [-0.3, 0.5]\, ,\\
        0.5 & \text{if } f_2(\omega) > 0.5 \, ,\\
        -0.3 & \text{if } f_2(\omega) < -0.3 \, .
        \end{cases} $$
All functions are evaluated on the same equispaced grid of $p=100$ points used to generate $\{ X_i(\omega)\}_{i=1}^n$. 

\begin{table}
\caption{Average Integrated Mean Squared Error (MISE) and its standard error (in parentheses) over 100 repetitions for the estimation of three regression functions (details in the text). TF-4: Trend filtering with penalization on fourth derivative only; TF-1: Trend filtering with penalization on the first derivative only; MTF: Trend filtering with penalization on both fourth and first derivative; SPL: Penalized splines as in \cite{goldsmith2011penalized}.}
\centering
\begin{tabular}{lccc}
\hline
\textit{Function} & $f_1(\omega)$ & $f_2(\omega)$ & $f_3(\omega)$ \\
\hline \\

\text{Scenario a)} & & & \\
\textit{TF-4} & 0.335 \footnotesize{(0.401)}  & 0.123 \footnotesize{(0.173)} & 0.265 \footnotesize{(0.073)}\\
\textit{TF-1} & 1.969 \footnotesize{(1.821)}  & 2.120 \footnotesize{(0.680)} & 0.595  \footnotesize{(0.339)}\\
\textit{MTF}  & 0.588 \footnotesize{(0.514)}  & 0.297 \footnotesize{(0.296)} & 0.184 \footnotesize{(0.081)}\\
\textit{SPL}  & 0.669 \footnotesize{(0.666)}  & 0.207 \footnotesize{(0.860)} & 0.357 \footnotesize{(0.405)}\\

\text{Scenario b)}  & & & \\
\textit{TF-4} & 0.382 \footnotesize{(0.434)}  & 0.139 \footnotesize{(0.189)} & 0.269 \footnotesize{(0.083)}\\
\textit{TF-1} &  1.584 \footnotesize{(0.335)} & 2.035 \footnotesize{(0.189)} & 0.561 \footnotesize{(0.069)}\\
\textit{MTF}  & 0.513 \footnotesize{(0.395)}  & 0.302 \footnotesize{(0.299)} & 0.194 \footnotesize{(0.083)}\\
\textit{SPL}  & 1.101 \footnotesize{(1.014)}  & 0.221 \footnotesize{(0.941)} & 0.368 \footnotesize{(0.477)}\\

\text{Scenario c)}  & & & \\
\textit{TF-4}  & 2.051 \footnotesize{(1.861} &  0.822  \footnotesize{(0.586)}  & 0.680 \footnotesize{(0.472)} \\
\textit{TF-1}  & 4.334 \footnotesize{(2.180)} & 2.705 \footnotesize{(1.091)} & 0.907 \footnotesize{(0.363)}  \\
\textit{MTF}   & 3.713 \footnotesize{(2.386)} & 1.165 \footnotesize{(0.974)} & 0.579  \footnotesize{(0.298)}  \\
\textit{SPL}   & 2.698 \footnotesize{(1.122)}  &  0.908 \footnotesize{(1.676)} & 0.630  \footnotesize{(0.366)} \\
\hline \\
\end{tabular}
\label{tab:est_err}
\end{table}

For all the $B = 100$ synthetic samples generated, we estimate the regression parameters with the trend filtering approach, penalizing the fourth derivative (\textit{TF-4}), the first derivative (\textit{TF-1}) and both of them (\textit{Mixed-TF}). For comparison purposes we also employ the spline method (\textit{SPL}) outlined in \cite{goldsmith2011penalized} penalizing the second derivative of the function. The tuning parameters for all the methods have been selected using a separate validation set.
In Table \ref{tab:est_err} we present for our methods and the spline estimator the value of the Integrated Mean Squared Error (MISE) defined as 
$$ \text{MISE}(\widehat{f}) = \int \{ f(\omega) - \widehat{f}(\omega) \}^2 d\omega,  $$ evaluated on the finite grid $\boldsymbol{\omega}$, averaged over all simulation repetitions, and its standard error (in parenthesis). The proposed approach shows superior performance in all the combinations of functions and scenarios considered, when compared to the spline methodology. In fact, this latter strategy is not well suited in situations where the regression function is spatially heterogeneous. Moreover we observe that, among the different specifications of the trend filtering, the one penalizing the fourth derivative achieves the best results in estimating $f_1(\omega)$ and $f_2(\omega)$, in all considered scenarios. Unfortunately, penalizing the first derivative does not lead to satisfactionary results, for two main reasons: the estimated regression function is not continuous, against what is commonly assumed in functional data analysis, and the estimation error is large due to inherent smoothness of the considered unknowns. However, adding the first derivative penalization to the plain trend filtering of order four leads to an improved performance if the regression function is particularly complex, as in the case of $f_3(\omega)$.
To elucidate the behaviour of double penalization in this scenario, in Figure \ref{fig:truncated} we graphically depict the unknown function and several estimates based on different values of the parameter ${\bm \lambda} = (\lambda_1, \lambda_2)$. Starting from the upper left corner where the impact of regularization is the lowest, we see that increasing $\lambda_1$ keeping  $\lambda_2$ fixed, leads to almost piecewise constant solutions. By contrast, increasing $\lambda_2$ while keeping $\lambda_1$ fixed, leads to almost piecewise cubic functions. However, since $f_3(\omega)$ exhibits both features, a better reconstruction is obtained by combining the two penalties, as can be appreciated in the lower right corner of the figure. Lastly, note that this specification automatically includes the ``marginal'' models with only one of the two derivative penalized.

%%%%%%%%%%%%%%%%%%%%%%%%%%%%%%%%%%%%%%%%%%%%%%%%%%%%

\section{Applications to milk spectral data}
\label{sec:application}

In the following, the proposed method is applied to the first set of data introduced in Section \ref{sec:data description}. Following suggestions from the literature, prior to running the analyses a variables aggregation step has been performed. In fact, it has been pointed out \citep[see e.g.,][]{murphy2010variable} that the aggregation of adjacent wavelengths implies almost negligible losses in terms of information and predictive abilities. This is coherent with the idea that, when dealing with spectra, the strong correlations among wavelengths allow to work on data with slightly lower resolution while retaining most of the informative content. Accordingly, we aggregate four adjacent wavelengths, to reduce the overall computational cost, resulting in a dataset with $n = 622$ milk spectra and $p = 264$ wavelengths.  

As briefly mentioned in Section \ref{sec:data description}, in the regression framework the proposed method has been used to predict the $\kappa$-casein content in the milk samples. The actual observed values for the response variable, expressed in grams per liter of milk, were collected using reverse-phase high performance liquid chromatography (HPLC), with an adaptation of the methodology considered in \citet{visser1991phenotyping}. This technology is known to be expensive and time-consuming and is not considered suitable for modern large-scale applications; therefore, the calibration of statistical tools, used in conjunction with infrared spectroscopy, can be highly beneficial for research in the dairy framework and for the dairy production systems. 

$\kappa$-casein has been selected as the milk trait to be predicted as it is one of the major components of milk, playing an essential role in cheese production systems, affecting both cheese yield and its characteristics \citep{wedholm2006effect}. Moreover, $\kappa$-casein is also used as a food additive and it generally represents an important economic factor whose timely and precise prediction might increase the efficiency of the dairy production chain. For these reasons, milk casein content is nowadays also considered as one of the determinants to estimate the breeding values of the animals, inspiring research lines on genetic control and selective breeding \citep{bittante2013genetic,bittante2022invited}. Exploratory analysis of this variable revealed a strong asymmetric behaviour of the empirical distribution. For this reason, we considered as a response variable for our model the logarithm of $\kappa$-casein. %\textcolor{violet}{Non vorrei dire una cazzata, ma credo che l'assunzione Gauss Markov sulla normalità sia sui residui, non sulle $y$. Magari possiamo dire che ci fa comodo il log per avere dominio $\mathbb{R}$.}

\begin{table}[t]
\centering
\caption{Estimated coefficients, and 95\% confidence intervals, for the scalar covariates.}
\label{tab:z_est}
\begin{tabular}{lrrr}
  \hline
Covariate & Lower ($0.025$) & Estimate & Upper ($0.975$)\\ 
  \hline
Intercept & 0.257 & $\boldsymbol{0.438}$ & 0.627 \\ 
  Spring & -0.222 & $\boldsymbol{-0.133}$ & -0.038 \\ 
  Summer & -0.075 & -0.019 & 0.026 \\ 
  Milk time (morning) & -0.186 & $\boldsymbol{-0.129}$ & -0.074 \\ 
  Parity (2) & -0.061 & -0.019 & 0.030 \\ 
  Parity (3) & -0.043 & 0.004 & 0.052 \\ 
  DIM & -0.001 & -0.000 & 0.000 \\ 
   \hline
\end{tabular}
\end{table}

\begin{figure}[t]
	\centering
	\begin{minipage}{\linewidth}
	    \centering
		%\textsf{(a) }\par\medskip
		\includegraphics[width=\linewidth]{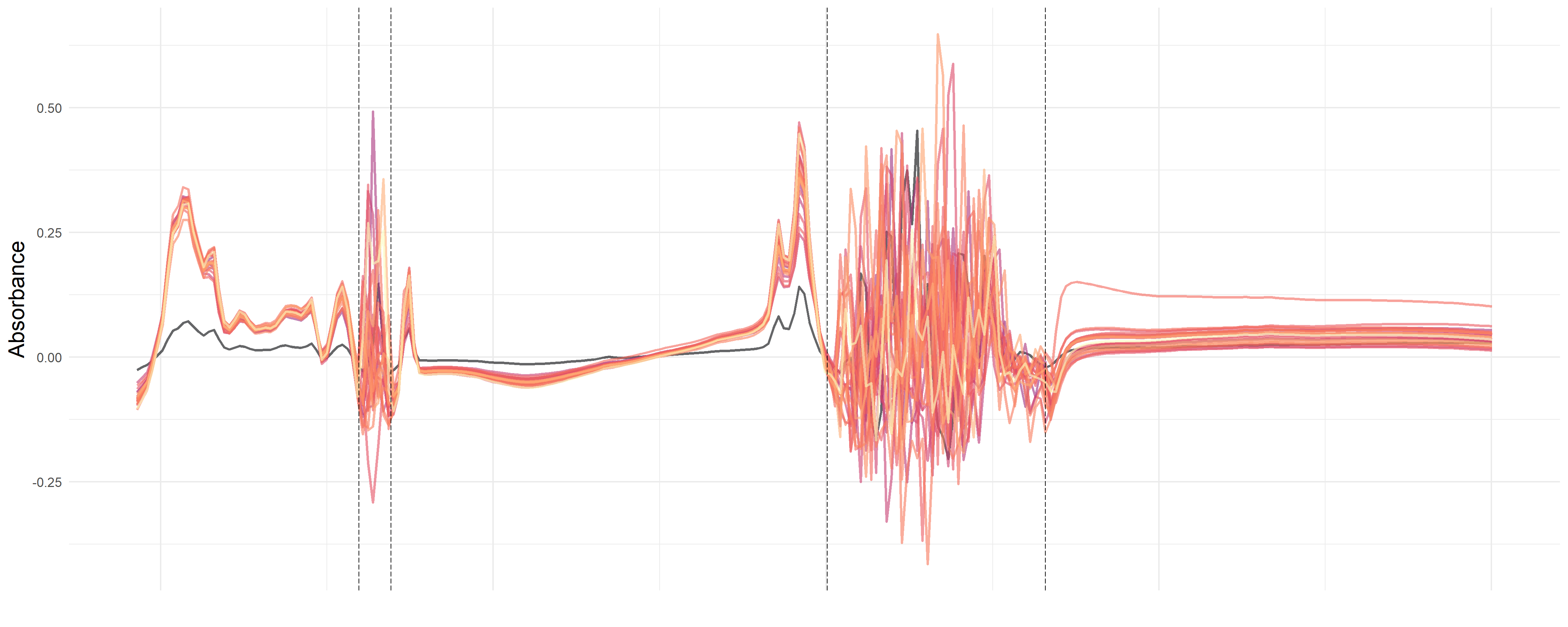}
		\end{minipage}
	\begin{minipage}{\linewidth}
		\centering
        \hspace{-0.2cm}  
		%\textsf{(b) }\par\medskip
		\includegraphics[width=\linewidth]{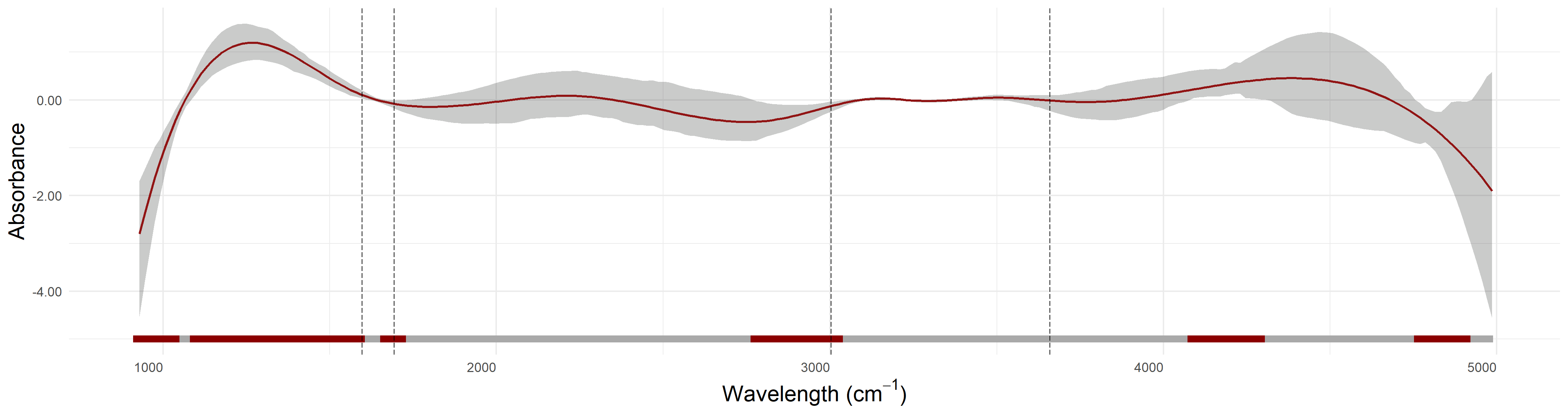}
		\end{minipage}
	\caption{Top:  Sample of 25 spectra, with dark (light) colors corresponding to low (high) values of $\kappa$-casein. Bottom: Estimated functional coefficient with $95\%$ bootstrap bands. The regions that do not contain zero are highlighted in red in the bottom line. }
	\label{fig:f_est}
\end{figure}

In this section, the model in \eqref{eq:problem4} has been considered, with $k = 3$ and $l = 0$, thus penalizing the fourth and the first derivative, respectively. This choice can be justified by the assumption of a  regression function that is smooth in some parts of the domain and flat in some others. Note, as mentioned in section \ref{sec:simstudy}, that the marginal formulations with penalization only on the fourth or on the first derivative are included as limiting models.
The hyperparameters $\lambda_1$ and $\lambda_2$, which control the strenght of the penalty, have been selected resorting to a cross-validation scheme. In conjunction with the spectral variables, we consider also some scalar covariates, as per the extension of the model outlined in Section \ref{sec:extensions}; in particular, information on the season when the milk samples have been collected, the milking time (morning or afternoon milking), the number of cows' parities and the number of days an animal has been milking in the current lactation (days in milk). This implies the presence of $r = 6$ additional scalar variables, with a total of 270 covariates. The estimated regression function, together with the inferential results obtained by means of the procedure introduced in Section \ref{sec:Inference}, are visually reported in Figure \ref{fig:f_est}, while the results concerning the scalar variables are shown in Table \ref{tab:z_est}. 

The method shows high prediction accuracy, with a cross-validated mean square prediction error equal to 0.04986. The result has been additionally compared with a PLS-based regression approach, unarguably representing the state-of-the-art when working with infrared spectroscopy data, which resulted in a cross-validated mean square prediction error of 0.05457.

As mentioned, our proposal introduces other relevant strength points while showing improved predictive performance. First, it respects and preserves the functional nature of the data, without mapping them into lower-dimensional latent spaces. Consequently, it provides richer insights on the studied phenomenon and, generally speaking, an easier interpretation of the results; this potentially sheds light on the chemical factors which are the main determinants of casein content in milk. In fact, a thorough analysis of the results depicted in Figure \ref{fig:f_est} highlights some interesting behaviours. First of all, the inferential routine outlined in Section \ref{sec:Inference}, allows to detect some spectral regions which are considered to be uninformative for the determination of the $\kappa$-casein content in the milk. For instance, our method considers as uninformative the spectral regions from 1619 cm$^{-1}$ to 1673 cm$^{-1}$ and from 3069 cm$^{-1}$ to 3663 cm$^{-1}$. In literature these highly-noisy regions are designated as water absorption areas, usually considered as uninformative, thus removed from the data prior to the analyses \citep{rutten2011prediction}. Nonetheless, the determination of these regions is still controversial and not unambiguous, as it can be influenced by spectrometer specific characteristics. Interestingly, our method marks as uninformative more wavelengths being adjacent to these highly noisy regions; this is coherent with practitioners' experiences which often point out that water may influence larger portions of the spectra, with respect to the ones suggested in the literature. 

Focusing on the variables regarded as significant, it has to be noted that the proposal suggests that $\kappa$-casein can be predicted using a relatively small portion of the infrared spectrum. This is consistent with the results obtained in \citet{FRIZZARIN20217438} where standard predictive tools displayed good performances exploiting fewer wavelengths, with respect to the ones used to predict other milk proteins and technological traits. These indications are particularly important for the dairy industry, where there is an increasing demand for cheaper and potentially portable instruments, scanning only relevant portions of the spectrum. 

A proper interpretation of the specific peaks shown in the estimated regression function is complex since, for composite materials as milk, chemical constituents have absorption peaks at different wavelengths which often overlap \citep{soyeurt2006estimating}. Nonetheless, some interesting behaviours can be highlighted. In general, we see a strong influence for the wavelengths in the so called \emph{fingerprint region}, below 1400 cm$^{-1}$ \citep{hewavitharana1997fourier}; this region is often regarded as highly informative for the analysis of proteinaceous material as here chemical bonds related to amide groups are formed \citep{van2002ftir}. Coherently, being $\kappa$-casein a protein, our method flags as influential, and with a positive effect on the $\kappa$-casein concentration, those wavelengths around 1550 cm$^{-1}$ and 1250 cm$^{-1}$ which are associated with amide II and amide III bands \citep{de2009prediction}. In the region around 1100 cm$^{-1}$ and between 1200 cm$^{-1}$ and 1300 cm$^{-1}$, the peaks often depend on the phosphate bands; interestingly, phosphorus in milk occurs only in casein and not in whey proteins \citep{hewavitharana1997fourier} and our method seems to be able to detect the importance of these areas. 

Lastly, some insights can be obtained by inspecting the results for the scalar covariates. For example, milk samples collected in spring appear to have a significant decrease in terms of $\kappa$-casein concentration; knowing that cows calve in the first months of the year, this is consistent with the suggestions in \citet{sorensen2003accuracy} where it is stated that casein concentration is usually lower after calving. Moreover, \citet{quist2008milking,forsback2010day} showed that casein content is higher for afternoon and evening milkings, with respect to the morning ones; as it can be seen in Table \ref{tab:z_est}, this is confirmed by our results.

Generally speaking, the devised procedure is capable of adequately predict the content of $\kappa$-casein in milk while, at the same time, paving the way for a convenient interpretation of the results, which is often more cumbersome when different predictive tools are considered. Finally, it should be noted that interpretation must be paired, as it might be enriched, by a close cooperation with an expert in the dairy and animal science framework.

\iffalse
\begin{figure}
    \centering
    \includegraphics[scale = 0.4]{estimation_functional_restricted4_10x10grid_new.pdf}
\end{figure}

\begin{figure}
    \centering
    \includegraphics[scale = 0.35]{estimation_scalar_restricted4_10x10grid_new.pdf}
\end{figure}
\fi

%%%%%%%%%%%%%%%%%%%%%%%%%%%%%%%%%%%%%%%%%%%

\subsection{Application to cow dietary treatments}

In this section, one of the extension discussed in Section \ref{sec:extensions}, is considered to analyze the second datasets introduced in Section \ref{sec:data description}. More specifically, assuming that the response variable arises from a Bernoulli distribution, we employ the proposed strategy to predict the cows' dietary treatment, relying only on the spectral information. Coherently with the application in the previous section, a wavelength aggregation step has been performed. Moreover, consistently with \citet{frizzarin2021application}, some outlying spectra have been removed. This results in a set of data with $n = 4261$ spectra, 2893 from pasture-fed cows and 1368 from TMR-fed ones, and $p = 264$ measured wavelengths. 

Hereafter, model \eqref{eq:problem3} has been employed; in particular we considered $k = 0$, therefore penalizing the first derivative, with the loss function adequately chosen to accomodate the binary nature of the response variable. The hyperparameter $\lambda$ has been selected again by cross-validation. The application of our method produces highly satistifactorily performances, resulting in a cross-validation missclassification error equal to 2.98\%. This result has been again compared with the one obtained by means of a PLS-based discriminant analysis strategy, which produced a similar cross-validated error equal to 2.60\%. This provides a strong indication about the suitability of our proposal, also when considered for classification purposes.  

Note that the extension to model \eqref{eq:problem3} of the procedure outlined in Section \ref{sec:Inference} is not trivial. Nevertheless, even if it is not possible to draw formal inferential conclusions on the estimated functional coefficient, a closer inspection of the result allows us to obtain relevant insights, which can be further explored and integrated with subject-matter knowledge. Firstly, the penalization on the first derivative allows one to obtain an estimated functional coefficient being flat and equal to zero, or having negligible magnitude, for the highly noisy spectral regions from 1604 cm$^{-1}$ to 1712 cm$^{-1}$, and from 3039 cm$^{-1}$ to 3810 cm$^{-1}$, which strongly overlaps with the water absorption areas, deemed irrelevant for discrimination. Small discrepancies with the results obtained in the previous section highlight how the proposed method could represent a completely data-driven and application-oriented way to detect uninformative spectral regions. Further inspection of the most relevant wavelengths leads to coherent indications, with respect to those available in the literature \citep[see e.g.,][]{frizzarin2021application}. For example, the \emph{fingerprint region} is again useful to discriminate between diets. Moreover, wavelengths between 2854 cm$^{-1}$ and 2977 cm$^{-1}$ seems to have a strong impact on the feeding regimens classification, thus agreeing with the suggestions in \citet{de2011effectiveness,lefevre2000interaction} where it is highlighted that this region is often used to estimated the milk fatty acid composition, which is in turn known to be highly correlated with the dietary treatment. 

Concluding the proposed classification tool, while respecting and preserving the functional nature of the data, is able to outperform state-of-the-art discriminative methods. Moreover, the inspection of the estimated coefficients allows us to gain relevant insights from a chemical standpoint, which might deserve further exploration from experts in the field. 

%%%%%%%%%%%%%%%%%%%%%%%%%%%%%%%%%%%%%%%%%%%%%%%%%%

\section{Conclusions and future directions}
\label{sec:conclusions}

In this work, we presented an adaptive functional framework for spectroscopy data, that stems from the trend filtering literature. The proposed regression method is characterized by high flexibility and adaptivity with respect to the complexity of the underlying data generating process. In particular, the method is capable of capturing different degrees of regularity of the slope function, while accounting for the high dimensionality and strong correlation among the wavelengths thanks to the $\ell_1$-regularization. The estimation is supported by a fast optimization procedure that leverages on the alternating direction method of multipliers framework, with a specialized acceleration scheme which provides superior convergence rates. The method is suitable for both Gaussian and non-Gaussian responses, and allows for the inclusion of scalar covariates, whose addition is often overlooked in the spectroscopy framework even if it might lead to better predictive performances. Moreover, the estimation strategy is enriched by a newly developed inferential procedure which allows to draw formal conclusions both on the functional and the scalar component. These are obtained with a nonparametric bootstrap approach, i.e. the wild bootstrap, that is particularly appropriate in high-dimensional regression models where the noise distribution is unknown.

The high adaptivity and the availability of inferential procedure are key features to enhance not only the interpretability of the results, but also their usability in real world scenarios. Indeed, spectroscopy data present peculiar statistical challenges, in particular intrinsic high-dimensionality of the inputs and strong correlation structures among the wavelengths. It is therefore paramount, from a practical perspective, to have a viable and interpretable tool that allows to carry out inference on specific regions of the spectrum, in order to gain relevant knowledge on specific properties of the samples (e.g. $\kappa$-casein content) or to highlight differences due to external factors (e.g. dietary treatments). 

The proposed methodology showed satisfactory performance in simulations and, more importantly, very promising results in the two spectroscopy-based data analyses. In terms of prediction accuracy, the results were either superior or comparable to the ones obtained by means of state-of-the-art techniques. In terms of inference, the flexibility of the model allowed a correct identification of the highly-noisy water absorption areas, without the necessity to remove such portions of the data prior to the analysis. Moreover, in both the regression and classification framework, informative peaks (e.g., those in the fingerprint region) were highlighted, providing interesting insights into which spectral regions affect certain properties of milk. The inclusion of covariates has also constituted a relevant advantage that resulted in interesting observations on the effect, for example, of the seasonality. It should be stressed that, even if the proposed methodology has been applied to MIR spectroscopy data, it may be extended to other data sharing similar features.

A first direction for future research might be the development of inferential procedures for the non-Gaussian response cases. This would
solidify the interpretability of the proposed methods even further, for instance in the classification framework. Moreover, this represents a particularly stimulating open problem, that might be approached via appropriate generalization of the nonparametric bootstrap procedures. Another possible extension could be the introduction of more complex penalties, that would allow the applicability of the method to a wider range of problems.

\section*{Acknowledgements}
This publication has emanated from research conducted with the financial support of Science Foundation Ireland (SFI) and the Department of Agriculture, Food and Marine on behalf of the Government of Ireland under grant number (16/RC/3835).

%--------------------------------------------------------------------------
\bibliographystyle{apalike}
\bibliography{biblio.bib}

\begin{thebibliography}{}

\bibitem[Alsberg, 1993]{alsberg1993representation}
Alsberg, B.~K. (1993).
\newblock Representation of spectra by continuous functions.
\newblock {\em Journal of Chemometrics}, 7(3):177--193.

\bibitem[Beer, 1852]{beer1852bestimmung}
Beer, A. (1852).
\newblock Bestimmung der absorption des rothen lichts in farbigen
  flussigkeiten.
\newblock {\em Annalen der Physik Chimie}, 162:78--88.

\bibitem[Berzaghi and Riovanto, 2009]{berzaghi2009near}
Berzaghi, P. and Riovanto, R. (2009).
\newblock Near infrared spectroscopy in animal science production: principles
  and applications.
\newblock {\em Italian Journal of Animal Science}, 8(sup3):39--62.

\bibitem[Bittante and Cecchinato, 2013]{bittante2013genetic}
Bittante, G. and Cecchinato, A. (2013).
\newblock Genetic analysis of the fourier-transform infrared spectra of bovine
  milk with emphasis on individual wavelengths related to specific chemical
  bonds.
\newblock {\em Journal of Dairy Science}, 96(9):5991--6006.

\bibitem[Bittante et~al., 2022]{bittante2022invited}
Bittante, G., Patel, N., Cecchinato, A., and Berzaghi, P. (2022).
\newblock Invited review: A comprehensive review of visible and near-infrared
  spectroscopy for predicting the chemical composition of cheese.
\newblock {\em Journal of Dairy Science}.

\bibitem[Boyd et~al., 2011]{boydADMM}
Boyd, S., Parikh, N., Chu, E., Peleato, B., and Eckstein, J. (2011).
\newblock Distributed optimization and statistical learning via the alternating
  direction method of multipliers.
\newblock {\em Foundations and Trends® in Machine Learning}, 3(1):1--122.

\bibitem[Casa et~al., 2022]{casa2021parsimonious}
Casa, A., O’Callaghan, T.~F., and Murphy, T.~B. (2022).
\newblock Parsimonious bayesian factor analysis for modelling latent structures
  in spectroscopy data.
\newblock {\em The Annals of Applied Statistics}, 16(4):2417--2436.

\bibitem[Codazzi et~al., 2022]{codazzi2022gaussian}
Codazzi, L., Colombi, A., Gianella, M., Argiento, R., Paci, L., and Pini, A.
  (2022).
\newblock Gaussian graphical modeling for spectrometric data analysis.
\newblock {\em Computational Statistics \& Data Analysis}, page 107416.

\bibitem[Crambes et~al., 2009]{funreg_splines}
Crambes, C., Kneip, A., and Sarda, P. (2009).
\newblock {Smoothing splines estimators for functional linear regression}.
\newblock {\em The Annals of Statistics}, 37(1):35 -- 72.

\bibitem[Davies and Kovac, 2001]{davies_kovac_algo}
Davies, P.~L. and Kovac, A. (2001).
\newblock {Local Extremes, Runs, Strings and Multiresolution}.
\newblock {\em The Annals of Statistics}, 29(1):1 -- 65.

\bibitem[De~Marchi et~al., 2009]{de2009prediction}
De~Marchi, M., Bonfatti, V., Cecchinato, A., Di~Martino, G., and Carnier, P.
  (2009).
\newblock Prediction of protein composition of individual cow milk using
  mid-infrared spectroscopy.
\newblock {\em Italian Journal of Animal Science}, 8(sup2):399--401.

\bibitem[De~Marchi et~al., 2011]{de2011effectiveness}
De~Marchi, M., Penasa, M., Cecchinato, A., Mele, M., Secchiari, P., and
  Bittante, G. (2011).
\newblock Effectiveness of mid-infrared spectroscopy to predict fatty acid
  composition of brown swiss bovine milk.
\newblock {\em Animal}, 5(10):1653--1658.

\bibitem[De~Marchi et~al., 2014]{de2014invited}
De~Marchi, M., Toffanin, V., Cassandro, M., and Penasa, M. (2014).
\newblock Invited review: Mid-infrared spectroscopy as phenotyping tool for
  milk traits.
\newblock {\em Journal of Dairy Science}, 97(3):1171--1186.

\bibitem[Dimatteo et~al., 2001]{rjmcmc_genovese}
Dimatteo, I., Genovese, C.~R., and Kass, R.~E. (2001).
\newblock {Bayesian curve‐fitting with free‐knot splines}.
\newblock {\em Biometrika}, 88(4):1055--1071.

\bibitem[Forsb{\"a}ck et~al., 2010]{forsback2010day}
Forsb{\"a}ck, L., Lindmark-M{\aa}nsson, H., Andr{\'e}n, A., {\AA}kerstedt, M.,
  Andr{\'e}e, L., and Svennersten-Sjaunja, K. (2010).
\newblock Day-to-day variation in milk yield and milk composition at the
  udder-quarter level.
\newblock {\em Journal of dairy science}, 93(8):3569--3577.

\bibitem[Frizzarin et~al., 2021a]{FRIZZARIN20217438}
Frizzarin, M., Gormley, I., Berry, D., Murphy, T., Casa, A., Lynch, A., and
  McParland, S. (2021a).
\newblock Predicting cow milk quality traits from routinely available milk
  spectra using statistical machine learning methods.
\newblock {\em Journal of Dairy Science}, 104(7):7438--7447.

\bibitem[Frizzarin et~al., 2021b]{frizzarin2021application}
Frizzarin, M., O'Callaghan, T.~F., Murphy, T.~B., Hennessy, D., and Casa, A.
  (2021b).
\newblock Application of machine-learning methods to milk mid-infrared spectra
  for discrimination of cow milk from pasture or total mixed ration diets.
\newblock {\em Journal of Dairy Science}, 104(12):12394--12402.

\bibitem[Goldsmith et~al., 2011]{goldsmith2011penalized}
Goldsmith, J., Bobb, J., Crainiceanu, C.~M., Caffo, B., and Reich, D. (2011).
\newblock Penalized functional regression.
\newblock {\em Journal of Computational and Graphical Statistics},
  20(4):830--851.

\bibitem[Hewavitharana and van Brakel, 1997]{hewavitharana1997fourier}
Hewavitharana, A.~K. and van Brakel, B. (1997).
\newblock Fourier transform infrared spectrometric method for the rapid
  determination of casein in raw milk.
\newblock {\em Analyst}, 122(7):701--704.

\bibitem[James, 2002]{james2002glm}
James, G.~M. (2002).
\newblock Generalized linear models with functional predictors.
\newblock {\em Journal of the Royal Statistical Society: Series B (Statistical
  Methodology)}, 64(3):411--432.

\bibitem[Johnson, 2013]{DP_johnson}
Johnson, N.~A. (2013).
\newblock A dynamic programming algorithm for the fused lasso and
  l$_0$-segmentation.
\newblock {\em Journal of Computational and Graphical Statistics},
  22(2):246--260.

\bibitem[Keller et~al., 2006]{keller2006infrared}
Keller, L.~P., Bajt, S., Baratta, G.~A., Borg, J., Bradley, J.~P., Brownlee,
  D.~E., Busemann, H., Brucato, J.~R., Burchell, M., Colangeli, L., et~al.
  (2006).
\newblock Infrared spectroscopy of comet 81p/wild 2 samples returned by
  stardust.
\newblock {\em Science}, 314(5806):1728--1731.

\bibitem[Kim et~al., 2009]{kim_TF}
Kim, S.-J., Koh, K., Boyd, S., and Gorinevsky, D. (2009).
\newblock $\ell_1$ trend filtering.
\newblock {\em SIAM Review}, 51(2):339--360.

\bibitem[Kong et~al., 2016]{kong2016partial}
Kong, D., Xue, K., Yao, F., and Zhang, H.~H. (2016).
\newblock {Partially functional linear regression in high dimensions}.
\newblock {\em Biometrika}, 103(1):147--159.

\bibitem[Lefevre and Subirade, 2000]{lefevre2000interaction}
Lefevre, T. and Subirade, M. (2000).
\newblock Interaction of $\beta$-lactoglobulin with phospholipid bilayers: a
  molecular level elucidation as revealed by infrared spectroscopy.
\newblock {\em International journal of biological macromolecules},
  28(1):59--67.

\bibitem[Mammen, 1993]{mammen1993bootstrap}
Mammen, E. (1993).
\newblock Bootstrap and wild bootstrap for high dimensional linear models.
\newblock {\em The Annals of Statistics}, 21(1):255--285.

\bibitem[McParland and Berry, 2016]{mcparland2016potential}
McParland, S. and Berry, D. (2016).
\newblock The potential of fourier transform infrared spectroscopy of milk
  samples to predict energy intake and efficiency in dairy cows.
\newblock {\em Journal of Dairy Science}, 99(5):4056--4070.

\bibitem[Morris, 2015]{review_morris}
Morris, J.~S. (2015).
\newblock Functional regression.
\newblock {\em Annual Review of Statistics and Its Application}, 2(1):321--359.

\bibitem[Morris et~al., 2008]{morris2008bayesian}
Morris, J.~S., Brown, P.~J., Herrick, R.~C., Baggerly, K.~A., and Coombes,
  K.~R. (2008).
\newblock Bayesian analysis of mass spectrometry proteomic data using
  wavelet-based functional mixed models.
\newblock {\em Biometrics}, 64(2):479--489.

\bibitem[Murphy et~al., 2010]{murphy2010variable}
Murphy, T.~B., Dean, N., and Raftery, A.~E. (2010).
\newblock Variable selection and updating in model-based discriminant analysis
  for high dimensional data with food authenticity applications.
\newblock {\em The Annals of Applied Statistics}, 4(1):396.

\bibitem[Müller and Stadtmüller, 2005]{gen_flm}
Müller, H.-G. and Stadtmüller, U. (2005).
\newblock {Generalized functional linear models}.
\newblock {\em The Annals of Statistics}, 33(2):774 -- 805.

\bibitem[O'Callaghan et~al., 2017]{o2017effect}
O'Callaghan, T.~F., Mannion, D.~T., Hennessy, D., McAuliffe, S., O'Sullivan,
  M.~G., Leeuwendaal, N., Beresford, T.~P., Dillon, P., Kilcawley, K.~N.,
  Sheehan, J.~J., et~al. (2017).
\newblock Effect of pasture versus indoor feeding systems on quality
  characteristics, nutritional composition, and sensory and volatile properties
  of full-fat cheddar cheese.
\newblock {\em Journal of Dairy Science}, 100(8):6053--6073.

\bibitem[O’Callaghan et~al., 2016a]{o2016quality}
O’Callaghan, T.~F., Faulkner, H., McAuliffe, S., O’Sullivan, M.~G.,
  Hennessy, D., Dillon, P., Kilcawley, K.~N., Stanton, C., and Ross, R.~P.
  (2016a).
\newblock Quality characteristics, chemical composition, and sensory properties
  of butter from cows on pasture versus indoor feeding systems.
\newblock {\em Journal of Dairy Science}, 99(12):9441--9460.

\bibitem[O’Callaghan et~al., 2016b]{o2016effect}
O’Callaghan, T.~F., Hennessy, D., McAuliffe, S., Kilcawley, K.~N.,
  O’Donovan, M., Dillon, P., Ross, R.~P., and Stanton, C. (2016b).
\newblock Effect of pasture versus indoor feeding systems on raw milk
  composition and quality over an entire lactation.
\newblock {\em Journal of Dairy Science}, 99(12):9424--9440.

\bibitem[Petrich, 2001]{petrich2001mid}
Petrich, W. (2001).
\newblock Mid-infrared and raman spectroscopy for medical diagnostics.
\newblock {\em Applied Spectroscopy Reviews}, 36(2-3):181--237.

\bibitem[Politsch et~al., 2020]{politsch2020trend}
Politsch, C.~A., Cisewski-Kehe, J., Croft, R.~A., and Wasserman, L. (2020).
\newblock Trend filtering--i. a modern statistical tool for time-domain
  astronomy and astronomical spectroscopy.
\newblock {\em Monthly Notices of the Royal Astronomical Society},
  492(3):4005--4018.

\bibitem[Porep et~al., 2015]{porep2015line}
Porep, J.~U., Kammerer, D.~R., and Carle, R. (2015).
\newblock On-line application of near infrared (nir) spectroscopy in food
  production.
\newblock {\em Trends in Food Science \& Technology}, 46(2):211--230.

\bibitem[Quist et~al., 2008]{quist2008milking}
Quist, M., LeBlanc, S., Hand, K., Lazenby, D., Miglior, F., and Kelton, D.
  (2008).
\newblock Milking-to-milking variability for milk yield, fat and protein
  percentage, and somatic cell count.
\newblock {\em Journal of Dairy Science}, 91(9):3412--3423.

\bibitem[Ramdas and Tibshirani, 2016]{ramdasADMM}
Ramdas, A. and Tibshirani, R.~J. (2016).
\newblock Fast and flexible admm algorithms for trend filtering.
\newblock {\em Journal of Computational and Graphical Statistics},
  25(3):839--858.

\bibitem[Ramsay and Silverman, 2005]{FDA1}
Ramsay, J.~O. and Silverman, B.~W. (2005).
\newblock {\em Functional data analysis}.
\newblock Springer, New York.

\bibitem[Reid et~al., 2006]{reid2006recent}
Reid, L.~M., O'donnell, C.~P., and Downey, G. (2006).
\newblock Recent technological advances for the determination of food
  authenticity.
\newblock {\em Trends in Food Science \& Technology}, 17(7):344--353.

\bibitem[Reiss and Ogden, 2007]{reiss2007functional}
Reiss, P.~T. and Ogden, R.~T. (2007).
\newblock Functional principal component regression and functional partial
  least squares.
\newblock {\em Journal of the American Statistical Association},
  102(479):984--996.

\bibitem[Rutten et~al., 2011]{rutten2011prediction}
Rutten, M., Bovenhuis, H., Heck, J., and van Arendonk, J. (2011).
\newblock Prediction of $\beta$-lactoglobulin genotypes based on milk fourier
  transform infrared spectra.
\newblock {\em Journal of dairy science}, 94(8):4183--4188.

\bibitem[Saeys et~al., 2008]{saeys2008potential}
Saeys, W., De~Ketelaere, B., and Darius, P. (2008).
\newblock Potential applications of functional data analysis in chemometrics.
\newblock {\em Journal of Chemometrics}, 22(5):335--344.

\bibitem[Shin, 2009]{partial_flm}
Shin, H. (2009).
\newblock Partial functional linear regression.
\newblock {\em Journal of Statistical Planning and Inference},
  139(10):3405--3418.

\bibitem[S{\o}rensen et~al., 2003]{sorensen2003accuracy}
S{\o}rensen, L.~K., Lund, M., and Juul, B. (2003).
\newblock Accuracy of fourier transform infrared spectrometry in determination
  of casein in dairy cows' milk.
\newblock {\em Journal of Dairy Research}, 70(4):445--452.

\bibitem[Soyeurt et~al., 2006]{soyeurt2006estimating}
Soyeurt, H., Dardenne, P., Dehareng, F., Lognay, G., Veselko, D., Marlier, M.,
  Bertozzi, C., Mayeres, P., and Gengler, N. (2006).
\newblock Estimating fatty acid content in cow milk using mid-infrared
  spectrometry.
\newblock {\em Journal of Dairy Science}, 89(9):3690--3695.

\bibitem[Talari et~al., 2017]{talari2017advances}
Talari, A. C.~S., Martinez, M. A.~G., Movasaghi, Z., Rehman, S., and Rehman,
  I.~U. (2017).
\newblock Advances in fourier transform infrared (ftir) spectroscopy of
  biological tissues.
\newblock {\em Applied Spectroscopy Reviews}, 52(5):456--506.

\bibitem[Tennyson, 2019]{tennyson2019astronomical}
Tennyson, J. (2019).
\newblock {\em Astronomical Spectroscopy: An Introduction to the Atomic and
  Molecular Physics of Astronomical Spectroscopy}.
\newblock World Scientific.

\bibitem[Tibshirani et~al., 2005]{fused_lasso}
Tibshirani, R., Saunders, M., Rosset, S., Zhu, J., and Knight, K. (2005).
\newblock Sparsity and smoothness via the fused lasso.
\newblock {\em Journal of the Royal Statistical Society: Series B (Statistical
  Methodology)}, 67(1):91--108.

\bibitem[Tibshirani, 2014]{trend_filtering}
Tibshirani, R.~J. (2014).
\newblock {Adaptive piecewise polynomial estimation via trend filtering}.
\newblock {\em The Annals of Statistics}, 42(1):285 -- 323.

\bibitem[Tibshirani and Taylor, 2011]{gen_lasso}
Tibshirani, R.~J. and Taylor, J. (2011).
\newblock {The solution path of the generalized lasso}.
\newblock {\em The Annals of Statistics}, 39(3):1335 -- 1371.

\bibitem[Van Der~Ven et~al., 2002]{van2002ftir}
Van Der~Ven, C., Muresan, S., Gruppen, H., De~Bont, D.~B., Merck, K.~B., and
  Voragen, A.~G. (2002).
\newblock Ftir spectra of whey and casein hydrolysates in relation to their
  functional properties.
\newblock {\em Journal of Agricultural and Food Chemistry}, 50(24):6943--6950.

\bibitem[Visentin et~al., 2015]{visentin2015prediction}
Visentin, G., McDermott, A., McParland, S., Berry, D.~P., Kenny, O., Brodkorb,
  A., Fenelon, M.~A., and De~Marchi, M. (2015).
\newblock Prediction of bovine milk technological traits from mid-infrared
  spectroscopy analysis in dairy cows.
\newblock {\em Journal of Dairy Science}, 98(9):6620--6629.

\bibitem[Visentin et~al., 2016]{visentin2016predictive}
Visentin, G., Penasa, M., Gottardo, P., Cassandro, M., and De~Marchi, M.
  (2016).
\newblock Predictive ability of mid-infrared spectroscopy for major mineral
  composition and coagulation traits of bovine milk by using the uninformative
  variable selection algorithm.
\newblock {\em Journal of Dairy Science}, 99(10):8137--8145.

\bibitem[Visser et~al., 1991]{visser1991phenotyping}
Visser, S., Slangen, C.~J., and Rollema, H.~S. (1991).
\newblock Phenotyping of bovine milk proteins by reversed-phase
  high-performance liquid chromatography.
\newblock {\em Journal of Chromatography A}, 548:361--370.

\bibitem[Wahba, 1990]{wahba_book}
Wahba, G. (1990).
\newblock {\em Spline Models for Observational Data}.
\newblock Society for Industrial and Applied Mathematics.

\bibitem[Wedholm et~al., 2006]{wedholm2006effect}
Wedholm, A., Larsen, L., Lindmark-M{\aa}nsson, H., Karlsson, A., and
  Andr{\'e}n, A. (2006).
\newblock Effect of protein composition on the cheese-making properties of milk
  from individual dairy cows.
\newblock {\em Journal of Dairy Science}, 89(9):3296--3305.

\bibitem[Yang et~al., 2016]{yang2016smoothing}
Yang, J., Zhu, H., Choi, T., and Cox, D.~D. (2016).
\newblock Smoothing and mean--covariance estimation of functional data with a
  bayesian hierarchical model.
\newblock {\em Bayesian Analysis}, 11(3):649.

\bibitem[Yao et~al., 2005]{funreg_fpca}
Yao, F., Müller, H.-G., and Wang, J.-L. (2005).
\newblock {Functional linear regression analysis for longitudinal data}.
\newblock {\em The Annals of Statistics}, 33(6):2873 -- 2903.

\bibitem[Zhao et~al., 2012]{zhao2012wavelet}
Zhao, Y., Ogden, R.~T., and Reiss, P.~T. (2012).
\newblock Wavelet-based lasso in functional linear regression.
\newblock {\em Journal of Computational and Graphical Statistics},
  21(3):600--617.

\bibitem[Zhou and Shen, 2001]{algo_knots}
Zhou, S. and Shen, X. (2001).
\newblock Spatially adaptive regression splines and accurate knot selection
  schemes.
\newblock {\em Journal of the American Statistical Association},
  96(453):247--259.

\end{thebibliography}

\end{document}